\PassOptionsToPackage{cmyk,table}{xcolor}
\documentclass[conference,flushend]{iaria} 
\pdfoutput=1 

\usepackage[babel=true,english=american]{csquotes}
\usepackage[USenglish]{babel}

\usepackage{amsmath}
\usepackage{enumerate}

\usepackage{biblatex}
\usepackage[%
	babel=true, 
	expansion=alltext,
	protrusion=alltext-nott, 
	nopatch=eqnum, 
	final 
]{microtype}

\title{Strategies for Intrusion Monitoring in Cloud Services}

\author{
    \IEEEauthorblockN{George R.~S. Weir}
    \IEEEauthorblockA{%
        Department of Computer and Information Sciences\\
        University of Strathclyde\\
        Glasgow, UK\\
        e-mail: {\tt george.weir@strath.ac.uk}
    }
    \and
    \IEEEauthorblockN{Andreas Aßmuth\,\orcidlink{0009-0002-2081-2455}}
    \IEEEauthorblockA{%
    	University of Applied Sciences\\
    	OTH Amberg-Weiden\\
    	Germany\\
    	e-mail: {\tt a.assmuth@oth-aw.de}
    }
}

\usepackage{ifpdf}
\ifpdf
\pdfoutput=1 
\pdftrue
\fi

\makeatletter
\def\ps@IEEEtitlepagestyle{
  \def\@oddfoot{\mycopyrightnotice}
  \def\@evenfoot{}
}
\def\mycopyrightnotice{
  {\footnotesize
    \begin{minipage}{0.8\textwidth}
    \centering
    Please cite as: \fullcite{selfref}.
    \end{minipage}
  }
}
\makeatother
\makeatletter
\let\blx@rerun@biber\relax
\makeatother
\usepackage[shortcuts]{extdash} 

\DeclareBibliographyCategory{selfref}
\addtocategory{selfref}{selfref}

\begin{document}
    
\maketitle

\begin{abstract}
Effective activity and event monitoring is an essential aspect of digital forensic readiness. Techniques for capturing log and other event data are familiar from conventional networked hosts and transfer directly to the Cloud context. In both contexts, a major concern is the risk that monitoring systems may be targeted and impaired by intruders seeking to conceal their illicit presence and activities. We outline an approach to intrusion monitoring that aims (i)~to ensure the credibility of log data and (ii)~provide a means of data sharing that supports log reconstruction in the event that one or more logging systems is maliciously impaired.
\end{abstract}


\section{Introduction}\label{sec:intro}
A news report from a recent computer electronics trade show featured a light bulb with an in-built spy camera. Although the application of this device is the realm of physical security rather than the world of computer, Clouds and networks, we can derive two general lessons from this example technology. Firstly, the purpose of the device is surveillance. Secondly, the device aims for covert operation. These joint concepts of covert surveillance are important in the context of security, whether in the home, on a network or in the Cloud. The primary role for this spying light bulb is surveillance, i.e., in the event of a security incident, to record data that may later have evidential value. Capturing such data in a covert manner aims to reduce the likelihood that the recording facility will be detected and thereby, minimise the prospect that the data collection will be deliberately impaired and the telling data subverted.\par 
While covert surveillance affords no immediate defence against security breaches, it does illustrate the desirability of establishing auditable data in order that light may later be shed on unauthorised or anomalous events that have initially gone undetected by relevant human agency. With varying degrees of transparency, the logging features in computer operating systems, individual computer applications, network operations and Cloud environments go some way toward addressing this requirement by recording data that may subsequently be consulted, in a process of digital forensics, as evidence of past events. Thereby, ‘a forensic investigation of digital evidence is commonly employed as a post-event response to a serious information security incident.’ \dots\par 
‘Forensic readiness is defined as the ability of an organisation to maximise its potential to use digital evidence whilst minimising the costs of an investigation’~\autocite[1]{rowlingson}.\par 
Although considerable efforts are directed in computer security toward protection and prevention of illicit access and system misuse, digital forensic readiness is increasingly recognised as a necessary measure toward recovery, understanding vulnerabilities and pursuit of those responsible for cyber-misdeeds (e.g.,~\autocite{reddy}).\par 
In the following, Section~2 reviews the characteristics of Cloud services and the facilities available to the customer. Section~3 characterises the attack context, with reference to recognised phases and the likely associated intruder behaviour. In Section~4, we elaborate upon the role of monitoring as a basis for forensic readiness in Cloud Services, with specific attention to the variety of strategies that may be employed, both overt and covert, as well as their likely effectiveness as mechanisms for event reconstruction and on-going resilience. Section~5 presents an example monitoring approach that contains specific aspects toward a solution to the forensic readiness problem in the Cloud context. As summarised in Section~6, our proposed approach would generate auditable information that can be used subsequently for digital forensics analysis in a post-hack scenario, within a setting of Cloud Services.

\section{Cloud Services}\label{sec:cloud-services}
In this section, we briefly review the characteristics of Cloud Services, in order to highlight the security concerns associated with different use contexts.\par 
The US National Institute of Standards and Technology (NIST), has provided a detailed account of Cloud Services~\autocite{mell}. This includes a description of typical service models:
\begin{itemize}
	\item Software as a Service (SaaS);
	\item Platform as a Service (PaaS); and
	\item Infrastructure as a Service (IaaS).
\end{itemize}
In the first case, the customer is given access to applications running on the service provider’s Cloud infrastructure, usually through a variety of client devices and software interfaces. Aside from specific application configuration options, in this arrangement the customer is given no control over the underlying Cloud infrastructure (op. cit., p.~2). This level of service extends from simple file storage, through hosted Web sites and database management to specific Web services, including RESTful applications~\autocite{shaikh}, and use of ‘containers’~\autocite{richardson}.\par 
In the second case, the customer is permitted to deploy their own applications on to the service provider’s Cloud infrastructure. Customer control extends to configuration and management of these Cloud-hosted applications but, as before, the customer has no facility to control any other aspects of the underlying Cloud infrastructure~\autocite[p. 2]{mell}.\par 
In the third case, the customer has greater scope for software deployment on to the Cloud infrastructure, extending to ‘arbitrary software, which can include operating systems and applications’ (op. cit.). Still, in this arrangement, the customer’s control is limited to the deployed software applications, including operating systems (e.g., virtual machines) and associated networking features (such as software firewalls)~\autocite[p. 3]{mell}.\par 
These service models characterise typical Cloud Service Provider (CSP) offerings and the increasing levels of access and software capability, is reflected in increasing levels of cost to the consumer. Notably, in each of these contexts, management and control of the Cloud infrastructure resides with the CSP, who must be relied upon to manage most security aspects that may impinge upon the purchased services.\par 
The range of applications and software facilities afforded by Cloud services is extensive, and indications are that many mission-critical services are moving to Cloud implementations as a means of limiting security concerns and assuring greater resilience. The virtual nature of Cloud services also means that system recovery or replacement can be quick, reliable and cost-effective (cf.~\autocite{benatallah}). Such outsourcing of local software applications is recognised as commercially attractive for factors, such as:
\begin{itemize}
	\item Cost (reduction in local expertise and local infrastructure);
	\item Reliability (service-level agreements can assure availability);
	\item Resilience (speedy recovery in the event of data or service loss);
	\item Technical extensibility (support for multiple instances of applications with increasing availability of service to meet growing demand).
\end{itemize}
To simplify the categories of Cloud uses, we may broadly differentiate two end-user contexts. In the first, the customer employs the Cloud service as a data storage facility. (This is a specific instance of the Software as a Service.) Here, security for the customer is limited to concerns of authorised access, continuity of service and data maintenance. In the second context, the end-user employs the Cloud service as a means of computation. (This broadly covers all other Cloud interaction.) Here, security for the customer extends to all traditional aspects, including data protection, access authentication, service misappropriation and service availability. While some of these issues may lie within the control of the consumer, the CSP has ultimate management of the infrastructure that affords all of the higher-level service provision. The extent to which the CSP can reliably manage the security and associated integrity of provided services, depends ultimately upon the availability of techniques for detecting and recording the details of any illicit operations that take place within the Cloud service context. Without recourse to such facilities, the CSP cannot be counted upon to maintain consumer services in a satisfactory fashion since there is lack of assurance that such services have not been infiltrated, impaired or subverted. In addition, ability for the CSP to restore services to pre-compromise level depends largely upon the CSP’s facility to identify any delta between pre- and post-intrusion services. Inevitably, this leads back to the issue of
digital forensic readiness as applied to the Cloud context.

\section{The Attack Context}\label{sec:attack-context}
In general, there are three phases to a successful cyber-attack:
\begin{enumerate}[1.]
    \item reconnaissance and information gathering;
    \item infiltration and escalation and, finally;
    \item exfiltration, assault and obfuscation.
\end{enumerate}
In phase~1, the adversary gathers any information needed to gain access to the system, e.g., open ports, versions of operating systems and software services, security measures (such as firewalls, IDS, etc.)~\autocite{rimal}. Using this information, the adversary gains access to the system in phase~2~\autocite{murphy}.\par 
The process of gaining access might consist of several steps, for example, if the adversary has to comprise another system first, in order to get into the actual target. In this process, the adversary also tries to escalate available privileges in order to gain super-user access to the system.\par 
In phase~3, the adversary extracts any information from the system that might prove to be useful~\autocite{andress}. If the goal of the attack is stealing confidential data, such as user accounts, passwords or credit card information, this data is extracted by the adversary and possibly sold to third parties. If the cyber-attack has another goal, e.g., sabotage, the adversary extracts the data needed to launch the actual assault, often triggered by a certain date or specific event. In any case, the adversary can be expected to perform whatever action is required to cover their tracks. Among other actions, they may install a rootkit that exchanges current files and services within the system with modified versions of these particular files and services. Such system modifications may extend to altering process information, e.g., a program to list all running processes on the system may be modified to list all running processes except for the processes run by the adversary. Additionally, the adversary may target existing log files that might contain traces of the intrusion.\par 
Such strategies are reflected in many network-based intrusions since, in many instances, network vulnerability is predicated upon known weaknesses in networked hosts.

\section{Monitoring Strategies}\label{sec:monitoring-strategies}
As previously noted, digital forensic readiness requires the monitoring and recording of events and activity that may impinge upon the integrity of the host system. Much of this capability is provided natively by the local system, using standardly available operating system logging, perhaps with additional active security monitoring, such as dynamic log analysis~\autocite{oliner} or key file signature monitoring~\autocite{kim}.\par 
The situation for Cloud-based services reflects in many respects the context of a networked host. Where a customer employs Cloud purely as a storage medium, minimum security requirements will seek to ensure authenticated access and secure data backup. In turn, the monitoring requirements associated with this service must capture details of user logins (including source IP, username and success or failure of login attempts). Additionally, any file operations that change the status of data stored under the account of that customer must also be recorded. In the event of unauthorised access (e.g., stolen user credentials), such default monitoring may offer little protection, aside from identifying the identity of the stolen credentials and recourse to subsequent backup data recovery. Such monitoring is essentially Operating System-based, albeit that in the Cloud setting, this OS may be virtual.\par 
This context of Cloud usage faces the same challenges in monitoring and security that confront any networked host, with the added complication that a Cloud-based virtual host may face added vulnerability via its hosting virtualiser~\autocite{reuben}. Furthermore, Cloud services are often configured to provide new virtual OS instances automatically to satisfy demand, and in turn, shut these down when demand falls. A side-effect of such service cycling is that system logs are lost to the customer, and subsequent digital forensic analysis may be unavailable.\par 
In the ‘traditional’ network setting, numerous techniques have been devised to afford post-event insight on system failures and unwelcome exploits. In all major operating system contexts, whether virtualised, Cloud-based or native, system logging affords the baseline for generating auditable records of system, network and user activity. Such system level monitoring is well understood and in the event of intrusion is likely to be a primary target in order to compromise the record and eliminate traces of illicit activity.\par 
For networked hosts and, by extension, as a monitoring strategy for local area networks, a wide-variety of Intrusion Detection Systems (IDS) have been developed and deployed with a view to rapid determination of malicious activity. These techniques may be rule-based (e.g.,~\autocite{ilgun}). In most cases, the IDS monitors and cross-correlates system-generated logs in order to identify anomalous event sequences. Many approaches to anomaly-based intrusion detection have been reported~\autocite{garcia,chapman,wang,zhou,patcha,sukhwani}. Inevitably, such systems may themselves become targets in order to inhibit their detection capability and maintain a ‘zero-footprint’ on the part of the intruder~\autocite{tedesco}.\par 
In a Cloud context, each node is using its own logging daemon or agent to log important events. But in comparison to a single computer, the log information might be essential and therefore relevant for the whole cloud infrastructure. For that reason, cloud infrastructures use a centralised log server that receives the log information of all attached nodes. The task of this log server is not only the recording of log files of all nodes but also to monitor the cloud infrastructure. In case of a cyber-attack, the log server ideally detects the attack (maybe assisted by an intrusion detection system) and starts countermeasures. This exposed role of the log server makes it a very attractive target for cyber-attacks itself, or, as described above, means that an adversary has to deal with the log server in phase 2. Since the hardware of such a log server might also break down even without any cyber-attack, in practice more than one log server is used at the same time to provide redundancy.\par 
A practical solution might consist of two log servers in "active-active-mode" which means that both are operating at the same time, but in case of one system failure, the other takes over for the whole cloud infrastructure. The operation of these two log servers might be supervised by a third server which in case of failure or attack sends an alarm to the administrator. Unfortunately, the problem stays more or less the same: this third monitoring server is a single point of failure and is therefore attractive as a target for any adversary attacking the cloud infrastructure. If an adversary manages to take out the monitoring server and to tamper with the log information on at least one of the two log servers, the Cloud provider might not be capable of determining which log files are correct and which are manipulated.\par 
Any logging service which is introduced in addition to the traditional daemons or agents has to meet at least the following constraints:
\begin{enumerate}
	\item the new logging service must not cause too much additional load, either on the nodes (concerning computation) or on the network (concerning network traffic), and;
	\item the computation of additional security measures in order to provide authenticity and integrity must be efficiently feasible. 
\end{enumerate}

\section{Example Monitoring Approach}\label{sec:ex-mon-approach}
Message Authentication Codes (MACs) as described in almost any textbook about cryptography can readily be used to address this monitoring dilemma. MACs can be constructed using cryptographic hash functions or using block ciphers, for instance. Either construction ensures efficient computation of the MACs under a secret key. MACs are used to provide authenticity and integrity; therefore, they meet both conditions.\par 
A solution that we propose starts with a secure boot process for each node of the Cloud infrastructure. During boot, the common log daemon or agent is started and it starts recording events in various log files. We suggest to compute a MAC for each event and to store these additional bits with the plaintext message of the event in the log file. We assume that the plaintext message also contains a time stamp. For the next event to be recorded in a log file, the plaintext of the event is concatenated with the previous MAC before computing the MAC for this event. This leads to a MAC chain which can be checked for each step using the plaintext and MAC of the previous event~- but only if the secret key is known. Since the adversary does not know the secret key, he is not capable of computing valid MACs and therefore not capable of tampering with the MAC chain in order to hide his tracks.\par 
The use of Message Authentication Codes is only the first step towards a solution to the problem. An adversary could simply delete or deliberately falsify all log files (including the MACs). This would probably make it impossible to reconstruct the steps of the cyber-attack in a post-hack analysis.\par 
In order to deal with this issue and to make use of the benefits of a Cloud infrastructure, we propose the additional step of using secret sharing techniques~- or so called threshold schemes~- as published by Adi Shamir in 1979~\autocite{shamir}.\par 
The idea is to divide some data $D$ into $n$ pieces $D_1,\dots, D_n$ in such a way that:
\begin{enumerate}[(a)]
	\item $D$ can be reconstructed easily of any $k<n$ pieces $D_i$
	\item the knowledge of only $k-1$ or even fewer pieces $D_i$ leaves the data completely undetermined.
\end{enumerate}
Shamir named such a scheme a "$(k, n)$ threshold scheme". He points out that by using such a $(k, n)$ threshold scheme with $n=2k-1$, it is necessary to have at least $k=\left\lceil\frac{n+1}{2}\right\rceil$ parts $D_i$ to reconstruct $D$. A lesser number of $\left\lfloor\frac{n}{2}\right\rfloor=k-1$ parts makes the reconstruction impossible.\par 
Shamir introduced a $(k, n)$ threshold scheme based upon polynomial interpolation. The data $D$ can be interpreted as a natural number and $p$ is a prime number with $D<p$. All of the following computations are made in the prime field $\operatorname{GF}(p)$. Given $k$ points in the 2-dimensional plane, $(x_1,y_1), \dots, (x_k, y_k)$ with distinct coordinates $x_i$, there is one and only one polynomial $q$ of degree $k-1$ such that $q(x_i)=y_i$ for all $i=1,\dots,k$. At first, the coefficients $a_1,\dots,a_{k-1}$ are chosen at random and $a_0=D$, which leads to the  polynomial
$$ q(x) = a_0 + a_1x + a_2x^2 + \dots + a_{k-1}x^{k-1}. $$
The $n$ different pieces of $D$ are computed as $D_1=q(1), D_i=q(i),\dots,D_n=q(n)$. Provided that their identifying indices are known, any subset of $k$ elements $D_i$ can be used to compute the coefficients $a_i$ of the polynomial $q$ which allow the computation of the data $D = q(0)$. From any subset of less or equal $k-1$ pieces $D_i$, neither the coefficients $a_i$ nor the data $D$ can be calculated. (For further details, we direct the reader to the original paper~\autocite{shamir}.)\par 
In our proposed solution to the problem of providing additional forensic information for post-hack analysis, $D$ is the data to be written in a log file: the plaintext message of the event, $n$ randomly chosen nodes of the cloud infrastructure and the corresponding MAC, computed from the concatenation of the event message, the previous MAC and the addresses of these $n$ nodes. The $n$ pieces $D_i$ that are derived from $D$ as stated before, and $D$ is sent to the traditional centralised log server. The $n$ pieces $D_i$ are additionally sent to the $n$ nodes which store this information. For the next event, we repeat this procedure but choose $n$ (possibly) different nodes.\par 
In case of a cyber-attack and if a post-hack analysis is necessary, at first all pieces of logging information are gathered from all nodes. Using the time stamps and the MAC chains, the order of the logged events can be reconstructed. The decentralised stored pieces of logging information are put together to reconstruct $D$ from any $k$ of the $n$ parts. This means, even if an adversary succeeds in manipulating some of the nodes and the centralised logging system, the events can be reconstructed. Finally, the integrity and authenticity of these events can be checked using the MAC chain.\par 
The proposed approach may identify and retain information on an intruder’s actions that result in stolen, modified or deleted data. This is a feature with growing importance, as legislative demands on data protection increase. For instance, the EU General Data Protection Regulation that is due to come into force in May 2018, will require companies to notify all breaches within 72~hours of occurrence, with a potential penalty of up to $4\,\%$ of global turnover based on the previous year's accounts.\par 
Note that this solution is not proposed as a general basis for monitoring the Cloud infrastructure. Rather, its purpose is to provide secure logging information for a post-hack analysis by distributing their parts randomly over all nodes. Thereby, reliable system monitoring can be established by means of multiple log servers, with the added assurance of Message Authentication Codes.

\section{Conclusion}\label{sec:concl}
Recognising the importance of securing log data as a basis for digital forensic reconstruction in the event of system intrusion, a multiple server solution combined with Message Authentication Codes affords a mechanism that allows for safe deposit and reconstruction of monitor data. This can operate in a Cloud setting in which each logging node is a virtual server.\par 
An important benefit from this distrusted solution is that digital forensic reconstructions are possible for virtual machines that are ‘cycled’, since their native OS logs can be maintained in a recoverable and verifiable form beyond the OS of those machines. This provides the safeguard of digital forensic readiness for Cloud customers in the event that an intruder accesses private data on the Cloud service and causes that system to cycle as an attempt to delete all traces of illicit data access.\par 
The possibility, however slight, that an intruder may gain access to and potentially compromise all peers in this configuration, can be mitigated by also allowing log data to transfer ‘upwards’ to one or more ‘superior’ systems (e.g., the parent operating systems in which the peer log servers are virtualised).\par 
Evidently, Cloud service provision has a requirement for robust monitoring that is sufficient to withstand direct assault from an intruder within the host context. Conventional OS monitoring goes some way toward providing the equivalent of a light bulb with an in-built spy camera, but needs to be enhanced with a reliable mechanism for validating and reconstituting log data, such as we have outlined in this paper.

\renewcommand*{\bibfont}{\footnotesize}
\setlength{\labelnumberwidth}{0.45cm}
\printbibliography[notcategory=selfref]

@INPROCEEDINGS{selfref,
      title={{Strategies for Intrusion Monitoring in Cloud Services}},
      author={George R. S. Weir and Andreas Aßmuth},
      booktitle={Proc of the 8th International Conference on Cloud Computing, GRIDs, and Virtualization (Cloud Computing 2017)},
      address = {Athens, Greece},
      month=feb,
      year={2017},
      _url = {https://thinkmind.org/index.php?view=instance&instance=CLOUD+COMPUTING+2017},
      issn = {2308\-/4294},
      _pages = {49--53}
    }

\end{document}